# Photonic integration of lithium niobate micro-ring resonators onto silicon nitride waveguide chips by transfer-printing

ZHIBO LI,[1,3] JACK A. SMITH,[2] MARK SCULLION,[1] NILS KOLJA WESSLING,[2] LOYD J. MCKNIGHT,[1] MARTIN D. DAWSON,[1,2] AND MICHAEL J. STRAIN [2,4]

[1]*Fraunhofer Centre for Applied Photonics, Technology and Innovation Centre, 99 George St., Glasgow, G1 1RD, UK*
[2]*Institute of Photonics, Dept. of Physics, University of Strathclyde, Technology and Innovation Centre, 99 George St., Glasgow G1 1RD, UK*
[3]*zhibo.li@fraunhofer.co.uk*
[4]*michael.strain@strath.ac.uk*

**Abstract:** The heterogeneous integration of lithium niobate photonic waveguide devices onto a silicon nitride waveguide platform via a transfer-printing approach has been demonstrated for the first time. A fabrication process was developed to make free-standing lithium niobate membrane devices compatible with back-end integration onto photonic integrated circuits. Micro-ring resonators in membrane format were lithographically defined by using laser direct writing and plasma dry etching. The lithium niobate micro-ring resonators were then transferred from their host substrate and released onto a silicon nitride waveguide chip. An all-pass ring resonator transmission spectrum was obtained in the 1.5 μm to 1.6 μm wavelength range, with a measured loaded Q-factor larger than $3.2 \times 10^4$.

**Introduction**

Advances in heterogeneous photonic materials integration are enabling the fabrication of photonic integrated circuits (PICs) with a range of optical functions not possible in monolithic single material systems. Key examples include the integration of active gain materials [1], solid state single photon emitters [2], and efficient photodetectors [3] on passive PICs. Non-linear optical materials with figures of merit significantly higher than established PIC platforms are important additions to the heterogeneous integration toolkit [4,5]. Amongst these, lithium niobate ($LiNbO_3$ or LN) is a particularly interesting option, possessing excellent optical properties, such as wide transparency window (400 nm to 5 μm), relatively high refractive indices ($n_o$: ~2.21, $n_e$: ~2.14 at 1.55 μm), strong linear electro-optic effect ($\gamma_{33}$: ~31 pm/V), and large third order non-linear optical coefficients [6]. LN been widely used in electro-optic modulators, broadband optical combs, wavelength converters and photon-pair sources, covering applications from classical optical communications to emerging quantum technologies [7-10]. Studies on LN date back to the 1960s but were typically limited to low refractive-index contrast and discrete optical components in bulk LN materials. Recent breakthroughs in the preparation of lithium niobate on insulator (LNOI) wafers and the processing of low loss LN waveguides by plasma etching has seen LNOI become a rapidly maturing platform for integrated optics. A growing range of experimental demonstrations of active and passive photonic devices have been achieved on this monolithic LNOI platform [11-15]. The availability of high quality LN thin film integrated optics opens up the potential for heterogeneous materials integration with foundry compatible silicon and silicon nitride platforms that remain the current standard in large scale PIC design.

In this paper we report a transfer printing based approach for the heterogeneous integration of pre-fabricated LN membrane waveguide devices onto a SiN platform. Shallow etched micro-ring resonators in LN are fabricated in a suspended geometry making use of a temporary resist anchor layer to enable deterministic membrane device release. LN resonators integrated on SiN waveguides demonstrate waveguide to waveguide coupling and measured loaded Q-factors of ~$3\times10^4$.

**Hybrid material design and mode confinement**

Some examples of LN heterogeneously integrated with Si or SiN platforms have been reported, with the majority being variations of wafer or die bonding of LN films onto waveguide devices [16-18]. Integration of a LN membrane layer onto a SiN waveguide platform has also been demonstrated using a transfer printing method [19]. There are two key factors in the design of compact LN devices on host PIC platforms. First, the optical mode field overlap with the LN material, $\Gamma_{LN}$, should be maximized for efficient non-linear optical performance. Second, to enable compact optical device geometries, such as ring resonators, waveguide bend losses should be minimized. In the schemes mentioned above, where the optical mode is guided predominantly by the waveguide layer in Si or SiN, there is a strong trade-off between the mode confinement to the LN and the achievable bend loss. Using a finite eigenmode solver we calculated bend losses as a function of radius for a range of different integrated LN devices, including thin-film-on-waveguide platforms, monolithic LNOI, and corresponding to this work, LN etched waveguide on SiN. The calculated bend losses and associated mode-overlap to the LN layer are presented in Fig.1. For consistency with the devices presented in this work a fundamental TE mode at 1577 nm was calculated for each model.

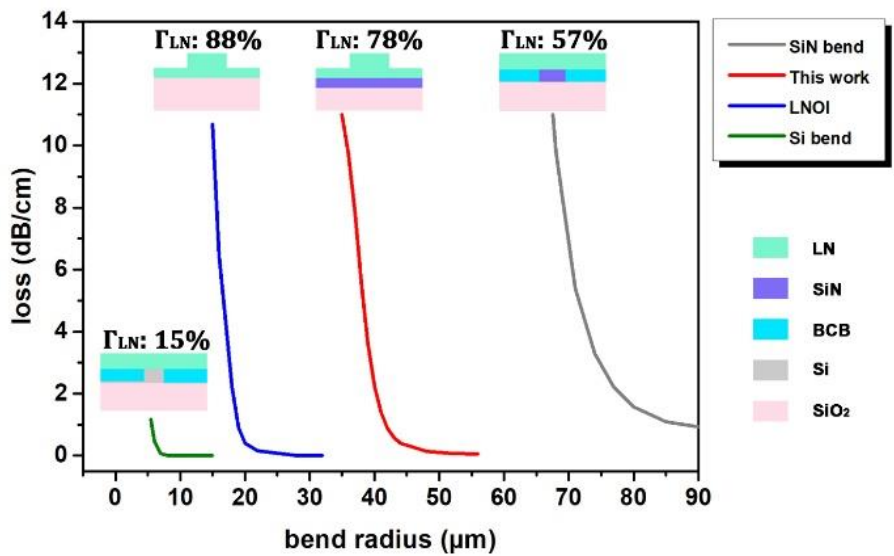

Fig.1, Simulated bend loss of the fundamental TE mode at 1577 nm as a function of bend radius for 4 designs of micro-ring resonator: Si micro-ring with a piece of LN bonded on top (green); monolithic LN micro-ring (blue); This work – LN micro-ring printed on SiN thin-film (red); SiN micro-ring with a bonded LN film (grey). The corresponding $|E|^2$ confinement in the LN layer ($\Gamma_{LN}$) is 15%, 88%, 78%, and 57%, respectively, calculated in the negligible bend loss region for each case. The schematic of each design is also included.

As expected due to the high refractive index of silicon, the lowest bend losses as a function of radius correspond to the LN-on-silicon device [20], which can support radii down to the few micron range, but also correspond to the lowest $\Gamma_{LN}$ of 15%. The highest $\Gamma_{LN}$ is presented by the monolithic LNOI platform, and since the optical mode is only guided in the LN layer, also shows low bend loss, highlighting the attractive properties of this as a single material platform.

The two options for LN-on-SiN show markedly different performance. By defining the waveguide structure in the LN thin film, the $\Gamma_{LN}$ is increased and the achievable bend losses are reduced in comparison with the case of an un-patterned LN film bonded onto a SiN waveguide. So, in general, for an effective hybrid LN-on-PIC platform, the index contrast should be in favour of the LN layer to ensure high $\Gamma_{LN}$ and the waveguide ridge should be defined in the LN to enable low bend losses.

**Device fabrication and integration**

Fig.2 presents the process flow developed to fabricate freestanding LN membrane micro-ring resonators for transfer print integration. A LNOI wafer, consisting of 600 nm x-cut LN device layer, 2 μm thermal silicon dioxide ($SiO_2$) and 500 μm Si substrate was used to make the membrane devices (material from NanoLN Ltd.). Fabrication started with the $SiO_2$ mask layer deposition (Fig.2a) by plasma enhanced chemical vapor deposition (PECVD). Micro-rings of 40 μm radius were defined in a photoresist layer by laser direct writing exposure (Heidelberg DWL66+) and transferred into $SiO_2$ by reactive ion etching (RIE). The rings were defined as a central 2.5 μm wide ridge in a shallow etched trench on both sides of the ring of 10 μm to avoid optical coupling from the waveguide to the surrounding slab regions. The LN was then etched 350 nm into the material by argon plasma in inductively coupled plasma (ICP) etching with $SiO_2$ as the hard mask (Fig.2b). Then a layer of $SiO_2$ was deposited (Fig.2c), and a second exposure and dry etching process were applied to define 100×100 μm² membrane mesa areas which encompassed the pre-fabricated micro-ring. Via-holes, to allow wet-etchant access to the sacrificial layer during membrane release, were also patterned in this lithography step (Fig.2d, corresponding microscope image in Fig.2g). It was found that the material strength of LN made it difficult to cleave commonly used monolithic tethers during the transfer printing process. The process detailed above results in a LN membrane design with no material tethers to the surrounding support material. Therefore, a sacrificial tether scheme was adopted to enable membrane release [23]. The final laser lithography step defined a membrane coating, with the same geometry and via-hole patterning as those of the LN membrane, together with trapezoid-shaped tethers in a 2 μm thick photoresist (Fig.2e, corresponding microscope image in Fig.2h). Finally, the sample was immersed in buffered hydrofluoric acid for the etch release of the $SiO_2$ under-cladding layer. After water rinse and nitrogen blow-dry, a suspended photoresist membrane together with the freestanding LN membrane micro-ring resonator had been fabricated (Fig.2f, corresponding microscope image in Fig.2i). This photoresist membrane acted as a supporting layer to increase the mechanical strength of the shallow etched ring resonator structure, thus avoiding LN membrane collapse or cleavage during the wet-etch and blow-drying steps. Interference fringe effects visible in Fig.2i are due to a small air gap between the membrane and substrate, indicating the membrane is in a suspended state.

The receiver chip based on the SiN photonic platform was fabricated in a 'Cornerstone' MPW (multi-project wafer) foundry run [24]. Waveguides were designed to operate under the single transverse-electric (TE) mode at the 1.55 μm waveband, with the SiN core of 300 nm in height and 1200 nm in width, and $SiO_2$ under-cladding of 3 μm in thickness. A 6 μm space gap was left on each side of the waveguide core to avoid optical coupling.

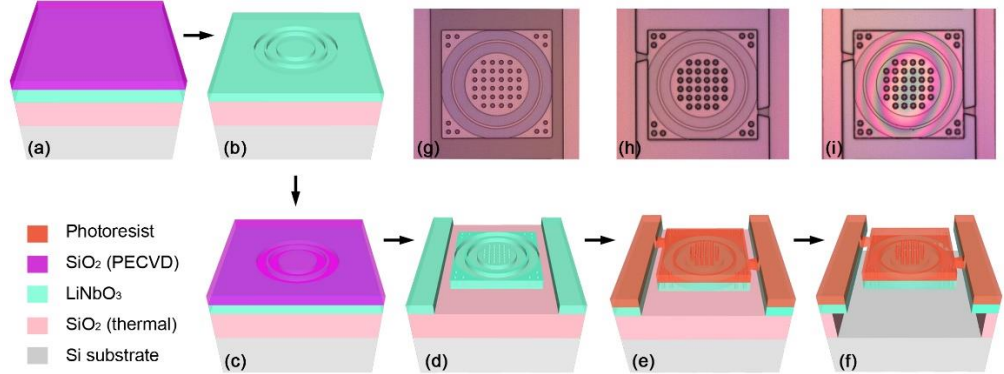

Fig.2. Process flow for the fabrication of freestanding LN membrane ring resonators, (a) SiO$_2$ deposition, (b) Microring on LNOI, (c) Second SiO$_2$ deposition, (d) LN membrane with via-holes but no tethers, (e) Photoresist membrane with via-holes and tethers on LN membrane, (f) Suspended photoresist membrane attached with freestanding LN membrane ring resonator. The optical microscopic images of (d), (e) and (f) are shown in (g), (h) and (i), respectively.

Fig.3 illustrates the transfer-printing procedure to assemble a LN membrane ring resonator onto the SiN waveguide platform, following the method detailed in [25]. An elastomeric polydimethylsiloxane (PDMS) stamp was used as a pick-up head and was first aligned to the targeted membrane device (Fig.3a). The stamp continued to move slowly toward the membrane until fully in contact, and a fast retraction caused the resist tethers to break (Fig.3b) [26]. The LN membrane was then transferred to a position above the receiver chip and aligned to the SiN waveguide with sub-micron accuracy [27]. As shown in Fig.3e, the square membrane was printed at an angle to the SiN bus waveguide to avoid optical point scattering from the un-etched corner portions of the LN membrane, allowing the SiN bus waveguide to traverse the shallow etched sections of the membrane. A slow retraction allowed the stamp to be released from the membrane, leaving the device bonded to the SiN waveguide chip (Fig.3c). Finally, the sample was cleaned in the oxygen plasma to remove photoresist residues (Fig.3d). The optical microscopic image of the finally assembled device is shown in Fig.3e, with an enlarged view showing the coupling region between LN ring resonator and SiN waveguide in Fig.3f. Due to the transparent nature of LN to visible light, the SiN waveguide is clearly visible underneath the LN membrane, and both outer edges of the waveguide and the ring were aligned.

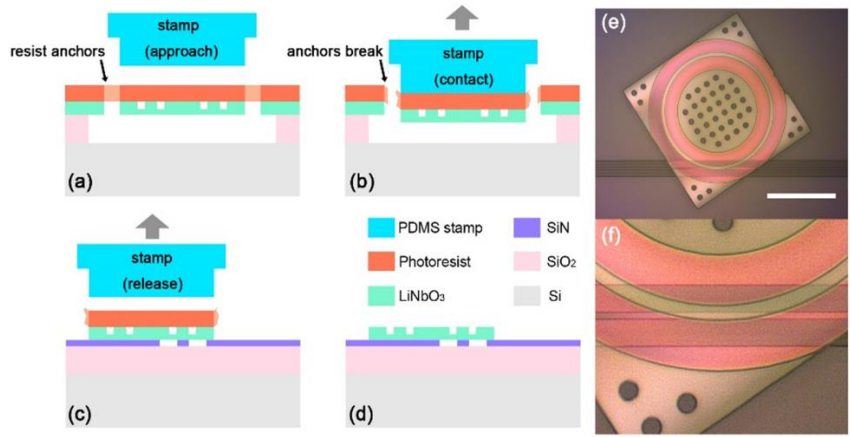

Fig.3. Diagram of transfer-printing, (a) stamp approaching to the LN membrane ring resonator, (b) contact of the membrane and breaking the resist anchor, (c) LN membrane printed on SiN waveguide, (d) finally assembled device after resist cleaning, (e) corresponding optical microscopic image, scale bar 50 μm, (f) enlarged view of the coupling region between SiN waveguide and LN ring resonator.

**Optical transmission results**

The final micro-assembled device was characterized using an end-fire laser optical transmission setup. Light generated by a tunable laser (Agilent 8164B) was coupled through a polarization filter into a polarization-maintaining single-mode lensed fiber for coupling into the SiN waveguide cleaved facet. TE polarization injection was set for this measurement. Light transmitted through the waveguide was collected by a microscope objective (20×, NA 0.3) and focused onto a free-space amplified photodetector (Thorlabs PDA20CS2). The measured optical spectrum was captured using an oscilloscope synchronized to the wavelength sweep of the tunable laser. To assess the propagation losses of the LNOI platform, a Fabry-Perot method [28] was used to measure monolithic LNOI waveguides (LNOI) fabricated with the same waveguide width and etch depth as the LN membrane micro-ring. Straight waveguide propagation losses of ~3 dB/cm were measured. These are higher than state-of-the-art LNOI waveguide devices in the literature [12] and are a result of residual waveguide roughness from our plasma etching process.

Fig.4 presents the optical transmission spectrum of the LN ring resonator printed on SiN waveguide, where the ring resonances have a free spectra range of ~4 nm.

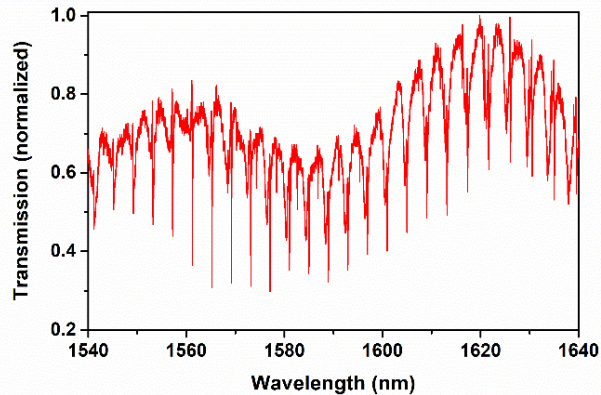

Fig.4, Normalized optical transmission spectrum of LN ring resonator hybrid integrated on SiN waveguide. Ring resonator: 40 µm in radius, 2.5 µm waveguide width, 350 nm etched depth.

Although only a single TE mode is supported in the SiN waveguide, the 2.5 μm wide LN ring resonator can also support high-order optical modes. The spectrum in Fig.4. exhibits three families of mode resonances. A selection of the spectrum around 1577 nm in Fig.5(a) shows these three resonances more clearly. A non-linear least-squares fit to an analytical model of an all-pass ring resonator was applied to each resonance to extract its coupling coefficient ($\kappa$), loaded and intrinsic Q-factors [29]. The two narrower linewidth resonance modes, at 1577.15 nm and 1578.48 nm respectively presented loaded (intrinsic) Q-factors of $3.2\times10^4$ ($3.9\times10^4$) and $2.6\times10^4$ ($2.7\times10^4$), as shown in Fig.5(c) and (d). Fig.5(b) shows the third resonance, a broad spectral dip at 1576.4 nm, corresponding to the existence of a lossy mode.

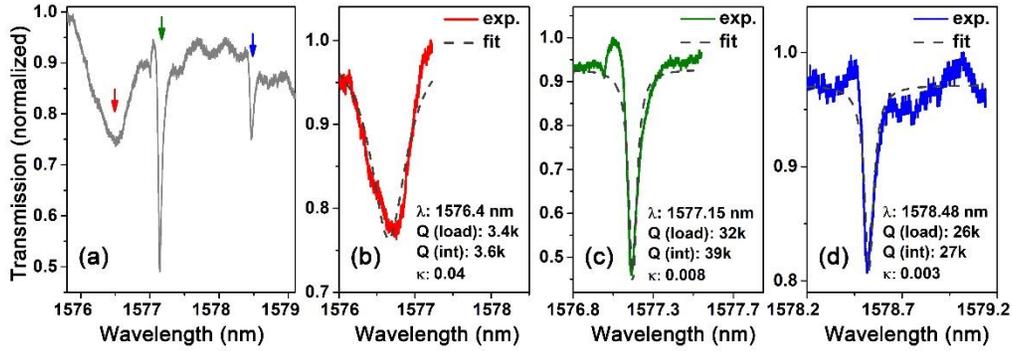

Fig.5, (a) Optical transmission spectrum at around 1577 nm, with arrows indicating three different resonant modes, the experimental and fitted resonance spectrum at 1576.4 nm, 1577.15 nm and 1578.48 nm is shown in (b), (c) and (d) respectively. Q (load): loaded Q-factor, Q (int): intrinsic Q-factor.

Numerical simulations were performed to visualize the cross-sectional optical mode distributions at the vertical coupling section between SiN waveguide and LN micro-ring resonator, by using a finite eigenmode solver in Lumerical MODE solutions. To reproduce the fabricated device, the edges of SiN waveguide and LN micro-ring were aligned in this model. Simulation results confirm that this hybrid LN micro-ring resonator supports three TE modes at 1577 nm, as shown in Fig.6. All three modes exhibit asymmetric distributions due to the non-centrosymmetric placement of the ring resonator relative to the waveguide. From the simulation, the TE02 mode shows significant electric field hybridization with ~83% TE polarization fraction. The asymmetric geometry of the micro-ring resonator and waveguide is likely to create some polarization rotation in the effective waveguide [30], thus exciting hybrid modes even with single TE mode input from the SiN waveguide.

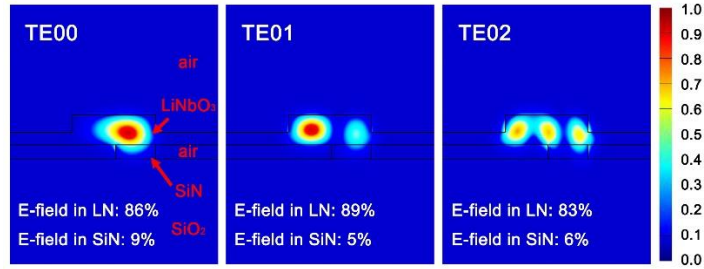

Fig.6, TE mode distributions at the coupling section of LN ring resonator printed on SiN waveguide, simulated wavelength of 1577 nm, SiN waveguide: height 300 nm, width 1200 nm; LN micro-ring: ridge height 350 nm, slab height 250 nm, width 2500 nm. E-field intensity was normalized, and the maximum intensity of hybrid TE02 mode was ~80% of those of TE01 and TE02 modes.

The highest loaded Q-factor from the measured device corresponds to the fundamental TE mode that exhibits the lowest bending losses of the three supported modes. The TE01 mode is still reasonably well confined, producing comparable loaded Q-factor values while the TE02 mode presents significant bend loss for the 40 μm radius ring and therefore modest Q-factor. The intrinsic Q-factor of the fundamental TE mode is $3.9 \times 10^4$, which although reasonable for modulator designs, is still significantly lower than state-of-the-art monolithic LNOI devices [12]. As noted previously, the additional losses in our demonstration are most likely due to sub-optimal plasma etching of the LN. Additionally, our bend radius of 40 μm is at the limit of the hybrid LN-on-SiN geometry for negligible bend loss and could be improved by either removing the SiN layer under the LN ring or by relaxing the bend radius. By improving the intrinsic waveguide losses as described, the performance of the heterogeneously integrated LN-on-SiN waveguide device presented here shows clear potential for the use of transfer printing integration for future hybrid materials systems.

**Conclusions**

In conclusion, we developed a fabrication process to produce freestanding LN membrane ring resonators and demonstrated the feasibility of transfer-printing of such LN waveguide resonators onto SiN waveguide chips. Our membrane fabrication process, without relying on any complex equipment such as critical point dryer or vapor etching, requires only one chemical wet etching step but effectively avoids the challenging issues of membrane collapse and cleavage due to capillary forces. In the current geometry the transmission spectrum and numerical simulations revealed that several optical modes were supported in the LN micro-ring resonator. The device exhibited a loaded Q-factor reaching $3.2 \times 10^4$. These developments enable the heterogeneous integration of LN membrane devices onto a non-native waveguide platform as a post-fabrication process.

**Acknowledgement**. This work was supported in part by the Engineering and Physical Sciences Research Council (EP/P013597/1, EP/R03480X/1, EP/V004859/1), Innovate UK (50414 QT Assemble), Fraunhofer UK and the Royal Academy of Engineering under the Research Chairs and Senior Research fellowships scheme. The CORNERSTONE project provided the foundry services for the fabrication of SiN waveguide chips under EPSRC funding (EP/T019697/1).

**Disclosures.** All authors declare no conflict of interest.

**References**

1. G. Roelkens, D. Van Thourhout, R. Baets, R. Nötzel, and M. Smit, "Laser emission and photodetection in an InP/InGaAsP layer integrated on and coupled to a Silicon-on-Insulator waveguide circuit," Optics express **14**, 8154-8159 (2006).
2. F. Peyskens, C. Chakraborty, M. Muneeb, D. Van Thourhout, and D. Englund, "Integration of single photon emitters in 2D layered materials with a silicon nitride photonic chip," Nature communications **10**, 1-7 (2019).
3. J. Brouckaert, G. Roelkens, D. Van Thourhout, and R. Baets, "Thin-film III-V photodetectors integrated on silicon-on-insulator photonic ICs," Journal of Lightwave Technology **25**, 1053-1060 (2007).
4. J. McPhillimy, S. May, C. Klitis, B. Guilhabert, M. D. Dawson, M. Sorel, and M. J. Strain, "Transfer printing of AlGaAs-on-SOI microdisk resonators for selective mode coupling and low-power nonlinear processes," Optics Letters **45**, 881-884 (2020).
5. W. Xie, C. Xiang, L. Chang, W. Jin, J. Peters, and J. E. Bowers, "Silicon-integrated nonlinear III-V photonics," Photonics Research **10**, 535-541 (2022).
6. D. Zhu, L. Shao, M. Yu, R. Cheng, B. Desiatov, C. Xin, Y. Hu, J. Holzgrafe, S. Ghosh, and A. Shams-Ansari, "Integrated photonics on thin-film lithium niobate," Advances in Optics and Photonics **13**, 242-352 (2021).
7. E. L. Wooten, K. M. Kissa, A. Yi-Yan, E. J. Murphy, D. A. Lafaw, P. F. Hallemeier, D. Maack, D. V. Attanasio, D. J. Fritz, and G. J. McBrien, "A review of lithium niobate modulators for fiber-optic communications systems," IEEE Journal of selected topics in Quantum Electronics **6**, 69-82 (2000).
8. M. Zhang, B. Buscaino, C. Wang, A. Shams-Ansari, C. Reimer, R. Zhu, J. M. Kahn, and M. Lončar, "Broadband electro-optic frequency comb generation in a lithium niobate microring resonator," Nature **568**, 373-377 (2019).
9. O. Alibart, V. D'Auria, M. De Micheli, F. Doutre, F. Kaiser, L. Labonté, T. Lunghi, É. Picholle, and S. Tanzilli, "Quantum photonics at telecom wavelengths based on lithium niobate waveguides," Journal of Optics **18**, 104001 (2016).
10. H. Jin, F. Liu, P. Xu, J. Xia, M. Zhong, Y. Yuan, J. Zhou, Y. Gong, W. Wang, and S. Zhu, "On-chip generation and manipulation of entangled photons based on reconfigurable lithium-niobate waveguide circuits," Physical review letters **113**, 103601 (2014).
11. M. Zhang, C. Wang, R. Cheng, A. Shams-Ansari, and M. Lončar, "Monolithic ultra-high-Q lithium niobate microring resonator," Optica **4**, 1536-1537 (2017).
12. C. Wang, M. Zhang, M. Yu, R. Zhu, H. Hu, and M. Loncar, "Monolithic lithium niobate photonic circuits for Kerr frequency comb generation and modulation," Nature communications **10**, 1-6 (2019).
13. D. Pohl, M. Reig Escalé, M. Madi, F. Kaufmann, P. Brotzer, A. Sergeyev, B. Guldimann, P. Giaccari, E. Alberti, and U. Meier, "An integrated broadband spectrometer on thin-film lithium niobate," Nature Photonics **14**, 24-29 (2020).

14. E. Lomonte, M. A. Wolff, F. Beutel, S. Ferrari, C. Schuck, W. H. Pernice, and F. Lenzini, "Single-photon detection and cryogenic reconfigurability in lithium niobate nanophotonic circuits," Nature communications **12**, 1-10 (2021).
15. X. Guo, L. Shao, L. He, K. Luke, J. Morgan, K. Sun, J. Gao, T.-C. Tzu, Y. Shen, and D. Chen, "High-performance modified uni-traveling carrier photodiode integrated on a thin-film lithium niobate platform," Photonics Research **10**, 1338-1343 (2022).
16. L. Chang, M. H. Pfeiffer, N. Volet, M. Zervas, J. D. Peters, C. L. Manganelli, E. J. Stanton, Y. Li, T. J. Kippenberg, and J. E. Bowers, "Heterogeneous integration of lithium niobate and silicon nitride waveguides for wafer-scale photonic integrated circuits on silicon," Optics letters **42**, 803-806 (2017).
17. P. O. Weigel, J. Zhao, K. Fang, H. Al-Rubaye, D. Trotter, D. Hood, J. Mudrick, C. Dallo, A. T. Pomerene, and A. L. Starbuck, "Bonded thin film lithium niobate modulator on a silicon photonics platform exceeding 100 GHz 3-dB electrical modulation bandwidth," Optics express **26**, 23728-23739 (2018).
18. M. He, M. Xu, Y. Ren, J. Jian, Z. Ruan, Y. Xu, S. Gao, S. Sun, X. Wen, and L. Zhou, "High-performance hybrid silicon and lithium niobate Mach–Zehnder modulators for 100 Gbit/s and beyond," Nature Photonics **13**, 359-364 (2019).
19. T. Vanackere, M. Billet, C. O. de Beeck, S. Poelman, G. Roelkens, S. Clemmen, and B. Kuyken, "Micro-transfer printing of lithium niobate on silicon nitride," in *2020 European Conference on Optical Communications (ECOC)*, (IEEE, 2020), 1-4.
20. L. Chen, Q. Xu, M. G. Wood, and R. M. Reano, "Hybrid silicon and lithium niobate electro-optical ring modulator," Optica 1, 112-118 (2014).
21. B. Corbett, R. Loi, W. Zhou, D. Liu, and Z. Ma, "Transfer print techniques for heterogeneous integration of photonic components," Progress in Quantum Electronics **52**, 1-17 (2017).
22. D. Jevtics, B. Guilhabert, A. Hurtado, M. Dawson, and M. Strain, "Deterministic integration of single nanowire devices with on-chip photonics and electronics," Progress in Quantum Electronics, 100394 (2022).
23. J. Zhang, G. Muliuk, J. Juvert, S. Kumari, J. Goyvaerts, B. Haq, C. O. d. Beeck, B. Kuyken, G. Morthier, D. V. Thourhout, R. Baets, G. Lepage, P. Verheyen, J. V. Campenhout, A. Gocalinska, J. O'Callaghan, E. Pelucchi, K. Thomas, B. Corbett, A. J. Trindade, and G. Roelkens, "III-V-on-Si photonic integrated circuits realized using micro-transfer-printing," APL Photonics **4**, 110803 (2019).
24. "https://www.cornerstone.sotonfab.co.uk/."
25. J. McPhillimy, B. Guilhabert, C. Klitis, M. D. Dawson, M. Sorel, and M. J. Strain, "High accuracy transfer printing of single-mode membrane silicon photonic devices," Optics Express **26**, 16679-16688 (2018).
26. M. A. Meitl, Z.-T. Zhu, V. Kumar, K. J. Lee, X. Feng, Y. Y. Huang, I. Adesida, R. G. Nuzzo, and J. A. Rogers, "Transfer printing by kinetic control of adhesion to an elastomeric stamp," Nature materials **5**, 33-38 (2006).
27. J. McPhillimy, D. Jevtics, B. J. E. Guilhabert, C. Klitis, A. Hurtado, M. Sorel, M. D. Dawson, and M. J. Strain, "Automated Nanoscale Absolute Accuracy Alignment System for Transfer Printing," ACS Applied Nano Materials **3**, 10326-10332 (2020).
28. T. Feuchter and C. Thirstrup, "High precision planar waveguide propagation loss measurement technique using a Fabry-Perot cavity," IEEE photonics technology letters 6, 1244-1247 (1994).
29. W. Bogaerts, P. De Heyn, T. Van Vaerenbergh, K. De Vos, S. Kumar Selvaraja, T. Claes, P. Dumon, P. Bienstman, D. Van Thourhout, and R. Baets, "Silicon microring resonators," Laser & Photonics Reviews **6**, 47-73 (2012).
30. Y. Wakabayashi, T. Hashimoto, J. Yamauchi, and H. Nakano, "Short waveguide polarization converter operating over a wide wavelength range," Journal of lightwave technology **31**, 1544-1550 (2013).